\documentclass[epj]{svjour}

\usepackage{graphicx}
\usepackage{color}
\usepackage{amsmath}
\usepackage{empheq}
\usepackage{hyperref}

\newcommand{\vecvar}[1]{\ensuremath{\mathbf{#1}}}
\newcommand{\funcder}[2]{\ensuremath{\frac{\delta#1}{\delta#2}}}
\newcommand{\pot}[1]{\ensuremath{\upsilon_{\text{#1}}}}
\newcommand{\potxc}{\ensuremath{\pot{xc}}}
\newcommand{\f}[1]{\ensuremath{f_{\text{#1}}}}
\newcommand{\fxc}{\ensuremath{\f{xc}}}
\newcommand{\vect}[1]{\ensuremath{\mathbf{#1}}}
\newcommand{\gradn}{\ensuremath{\nabla n(\vect{r})}}
\newcommand{\gradd}{\ensuremath{\nabla \delta(\vect{r}-\vect{r'})}}

\newcommand{\re}{\ensuremath{\vect{r}}}
\makeatletter
\renewcommand\@maketitle{\newpage
 \normalfont
 \setbox\authrun=\vbox\bgroup
 {\Large \bfseries\boldmath
  \noindent\ignorespaces
  \@title \par}\vskip 11.24pt\relax
 \if!\@subtitle!\else
   {\large\bfseries\boldmath
   \pretolerance=10000
   \rightskip=0pt plus 3cm
   \noindent\ignorespaces\@subtitle \par}\vskip 11.24pt
 \fi
 \normalfont\authorfont
 \lineskip .5em
 \setbox0=\vbox{\setcounter{auth}{1}\def\and{\stepcounter{auth} }%
                \hfuzz=2\textwidth\def\thanks##1{}\@author}%
 \value{inst}=\value{auth}%
 \setcounter{auth}{1}%
 \rightskip=0pt plus 2cm
 \noindent\ignorespaces\@author\vskip7.23pt
 \rightskip=0pt\relax
 \normalfont\small\rmfamily
 \institutename
 \vskip 12.85pt \noindent\@date
 \if!\@dedic!\else
    \par
    \small\itshape
    \addvspace\baselineskip
    \noindent\@dedic
 \fi
 \egroup 
 \vskip-30pt
 \unvbox\authrun
 \global\@minipagetrue
 \global\everypar{\global\@minipagefalse\global\everypar{}}
 \vskip22.47pt
}
\makeatother
\begin{document}
\title{On the challenge to improve the density response with unusual gradient approximations}
\author{J.\ Garhammer\inst{1} \and F. \ Hofmann\inst{1} \and R. Armiento\inst{2} \and  S. K\"ummel\inst{1}
}                     
\institute{Theoretical Physics IV, University of Bayreuth, 95440 Bayreuth, Germany \and Link\"oping University, Dept.\ of Physics, Chemistry and Biology, SE-581 83 Link\"oping, Sweden}
\date{\textbf{Preprint} May 25, 2018\\\strut\\
\textbf{Abstract.} Certain excitations, especially ones of long-range charge transfer character, are poorly described by time-dependent density functional theory (TDDFT) when typical (semi-)local functionals are used. A proper description of these excitations would require an exchange-correlation response differing substantially from the usual (semi-)local one. It has recently been shown that functionals of the generalized gradient approximation (GGA) type can yield unusual potentials, mimicking features of the exact exchange derivative discontinuity and showing divergences on orbital nodal surfaces. We here investigate whether these unusual potential properties translate into beneficial response properties. Using the Sternheimer formalism we closely investigate the response obtained with the 2013 exchange approximation by Armiento and K\"ummel (AK13) and the 1988 exchange approximation by Becke (B88), both of which show divergences on orbital nodal planes. Numerical calculations for $\mathrm{Na}_{2}$ 
as well as analytical and numerical calculations for the hydrogen atom show that the response of AK13 behaves qualitatively different from usual semi-local functionals. However, the AK13 functional leads to fundamental instabilities in the asymptotic region that prevent its practical application in TDDFT. Our findings may help the development of future improved functionals, and corroborate that the frequency-dependent Sternheimer formalism is excellently suited for running and analyzing TDDFT calculations.
} 

\maketitle

\section{Introduction}

Kohn-Sham (KS) density functional theory (DFT)~\cite{hk,ks} and its time-dependent extension (TDDFT) by Runge and Gross~\cite{RG84} are highly successful and among the most widely used theoretical approaches for describing the electronic structure and dynamics in physical, chemical and biological systems. Many applications of TDDFT are concerned with predicting the linear response. Consequently, the linear response of the exchange-correlation (xc) potential to a time-dependent perturbation, which has been studied in detail by E.K.U.\ Gross~\cite{grosskohnfxc,petersilka,lein2000,grabo2000}, to whom this special issue is devoted, plays a prominent role in TDDFT research. Commonly used functionals such as the local-density approximation (LDA), usual generalized gradient corrections (GGAs) such as the one of Perdew, Burke and Ernzerhof~\cite{pbe} (PBE), and hybrid functionals~\cite{B93b} predict many properties reliably. At the same time, however, they are known to systematically fail for certain problems. One such prominent failure of (semi-)local functionals and usual hybrid functionals with moderate fractions of exact exchange is their qualitatively wrong prediction of long-range charge-transfer phenomena~\cite{dreuw04,tozer03,maitraCT,ctreview2017}.

In recent years, semi-local exchange functionals and model-potentials have been developed which yield physically interpretable eigenvalues~\cite{tozer2015} and show features in their potentials that are very similar to important features of the exact Kohn-Sham exchange (EXX) potential. Prominent examples of this development are the Becke-Johnson model potential~\cite{bj} with its different modifications~\cite{gaidukjcp08,staroverov08,AKK08,KAK09,RPP10,koller2012}, especially the Tran-Blaha model potential~\cite{tran2009,tran_band_2007,tran2015}, and Becke-Johnson inspired new developments such as the AK13 functional~\cite{ak13}. A considerable part of the great interest in these developments stems from the hope that such functionals may allow to obtain information about excited states and the density response accurately at moderate computational cost~\cite{marquesak13,karolewskiPRA}. We review the corresponding arguments in detail in the next section. However, the Becke-Johnson potential cannot be used reliably in TDDFT calculations~\cite{karolewskiPRA}, because it is not a functional derivative~\cite{KAK09,stray}. As a consequence, TDDFT calculations with the Becke-Johnson model potential in general will be unstable, e.g., due to zero-force theorem violations~\cite{mundtzerof}. Similar conclusions hold for other model potentials. 

 Hence, the focus of the present work is a careful investigation of the response of the AK13 exchange energy functional, which shares many features with the Becke-Johnson model, yet is a functional dervative. In the present context, the most important feature that the AK13 and Becke-Johnson potential have in common is that for a finite system, the potential asymptotically goes to a value that is determined by the highest occupied eigenvalue. This leads to a discontinuity-like potential step structure similar to exact exchange. As such discontinuities are important for charge-transfer excitations~\cite{tozer03}, one may hope that a potential with such features may lead to a proper description of those. Therefore, our present study of the TDDFT response of the AK13 functional is, to the best of our knowledge, the first investigation of whether such semi-local step structures have a benefical impact in TDDFT calculations, and in how far the concept of a ``potential with a non-vanishing asymptotic constant'' is beneficial in TDDFT.

Summarizing our findings, we have to note that on the one hand, the answer that our study gives is largely negative: The AK13 response yields instabilities in the asymptotic region that prevent its use in TDDFT. On the other hand, the outcome clearly demonstrates that semi-local functionals designed to mimic exact exchange potential features are capable of giving a response that deviates strongly from the one that is observed with usual semi-local functionals. Thus, our results motivate future work on semi-local functionals that achieve an improved response, yet circumvent instabilities. We also expect the methodology and in-depth analysis presented in this work to be useful for future work in the area of designing functionals with improved response properties. Furthermore, our comparison between analytical and numerical results also adds confidence in the ability of the Sternheimer linear response formalism to correctly describe the response of difficult potentials.

The paper is organized as follows. We first review properties of exact and approximate exchange in TDDFT that are of particular relevance for excitations and thus motivate our study of the AK13 response in detail. Next we briefly review the functionals that we test, followed by a recapitulation of the Sternheimer linear response formalism that we use for our TDDFT calculations. After this we present numerical calculations for the sodium dimer as a simple test system. In order to explain the numerical findings that emerge, we then go through the analytically solvable case of the one-electron atom. We close by drawing our conclusions and offering an outlook.

\section{Exchange response in DFT and TDDFT}

Fock exchange is very frequently employed in DFT as a part of hybrid functional constructions and ameliorates deficiencies of usual (semi-)local approximations, e.g., by providing some non-locality to the functional and by reducing self-interaction errors~\cite{ksvgks}. However, the use of Fock exchange comes at a twofold price. On the practical side, the computational expense of exchange integrals is a burden. On the fundamental side, it has been argued since the beginnings of modern DFT (see, e.g., Ref.~\cite{williams1983} for examples) that it may be more consistent with the intrinsic many-body nature of DFT to approximate exchange and correlation together rather than dividing into single-particle motivated exchange and Coulomb correlation.

While the use of Fock exchange has proven beneficial in ground-state calculations, as testified by the success of hybrid functionals in questions of thermochemistry~\cite{B93}, many of the advantages of using Fock exchange as part of density functional approximations are not related to ground-state observables, but to the use of such functionals in TDDFT. Furthermore, some of the interest in the Becke-Johnson and related approximations has originated from the description of excitations~\cite{marquesak13,karolewskiPRA}. One can readily understand why Fock exchange can be beneficial in TDDFT from arguments based on linear response theory: following, e.g., Refs.~\cite{petersilka,casida96,tddftbook2006p10} one can interpret the true excitations as resulting from a combination of Kohn-Sham eigenvalue differences and exchange-correlation (xc) kernel corrections via matrix-elements of the type
\begin{equation}
K_{ijkl}=\int \int \varphi_i^*(\vect{r})\varphi_j(\vect{r})\frac{\delta v_\mathrm{xc}(\vect{r},t)}{\delta n(\vect{r}',t')} \varphi_k(\vect{r}') \varphi_l^*(\vect{r}') \, d^3r\, d^3r' \, ,
\label{eqn:casidaelement}
\end{equation} 
where spin indices have been suppressed for clarity of notation. From this perspective, two advantages of using Fock exchange in TDDFT become obvious. First, it typically leads to an eigenvalue spectrum of greater physical interpretability, and this can translate into improved TDDFT excitation energies~\cite{petersilka,filippi,vanmeer2014,fabian}. Second, step structures of the EXX potential~\cite{kli92steps,mundtprl,hellgrengross} or xc potential~\cite{leinkum,adiaexPRL,maitraCT,maitra2014,maitra2016} can translate into substantial effects in eq.~(\ref{eqn:casidaelement}), leading to large and important corrections to the single-particle eigenvalue differences. 

The latter argument is at the heart of understanding one of the most notorious failures of TDDFT with usual (semi-)local functionals, viz.\ its massive underestimation of long-range charge transfer excitation energies~\cite{dreuw04,tozer03}: As argued, e.g., in Refs.~\cite{dreuw04,maitraCT,ctreview2017}, long-range charge-transfer excitations correspond to situations where the orbital overlap in eq.~(\ref{eqn:casidaelement}) is small, vanishing exponentially as a consequence of exponential orbital decay. Thus, the matrix elements of eq.~(\ref{eqn:casidaelement}) vanish unless $\delta v_\mathrm{xc}(\vect{r},t)/\delta n(\vect{r}',t')$ counters the exponential orbital decay. When Fock exchange is used, the vanishing orbital overlap does not lead to vanishing $K_{ijkl}$ because EXX (and also an exact calculation including correlation~\cite{exactkernel}) leads to a non-local kernel, i.e., a kernel that also couples regions of space in which $\vect{r}$ and $\vect{r}'$ are far apart. The kernel of (semi-)local functionals, however, is local, i.e., $\propto \delta(\vect{r}-\vect{r}')$. Therefore, $K_{ijkl}$ will vanish for vanishing orbital overlap, erroneously making the TDDFT excitation energy equal to the Kohn-Sham eigenvalue difference, unless the spatial dependence of $v_\mathrm{xc}(\vect{r})$ is such that $\delta v_\mathrm{xc}(\vect{r},t)/\delta n(\vect{r},t')$ itself grows rapidly in regions of space in which the orbital overlap vanishes.

The potential of the LDA and usual GGAs follow the density closely. Therefore, they do not show a rapid increase or divergence of the kernel in regions of vanishing orbital overlap. Consequently, these approximations fail utterly in the description of long-range charge-transfer excitations~\cite{dreuw04,tozer03,ctreview2017,steinjacs}. As this physics is decisive in many highly-relevant questions of material science, with solar-cell development being a prominent example~\cite{karolewskiPCCP2013,korzdorfer2014,Li2015,Li2016,Li2016_2}, (semi-)local functionals to date are of only very limited use in this type of research. 

For a long time, it had been believed that closely following the density is an unavoidable feature of \mbox{(semi-)local} approximations. However, it recently has been demonstrated that 
a functional of the GGA type can have a functional derivative, i.e., a corresponding potential, that resembles exact exchange in several ways~\cite{tran2015,ak13,marquesak13,ak13bands,tran2016}. The hope that this functional can be widely used in ground-state material science calculations has been curbed by the yet more recent discovery~\cite{thilo1,thilo2} that it, and several other constructions following a related logic~\cite{b88,lb94,lembarki}, show divergences in regions of space where the highest occupied molecular orbital (HOMO) has a nodal plane overlapping with a lower occupied orbital. While this feature makes ground-state calculations difficult, it appears attractive from the perspective of TDDFT, where pronounced features of the potential in regions of reduced orbital overlap are required as discussed above.

Therefore, we calculate and analyze the linear response of such semi-local approximations in this manuscript. In order to circumvent issues resulting from the previously reported possible difficulties in ground-state calculations with such functionals, and in order to focus on and bring out the effects of the xc corrections as clearly as possible, we resort to the Sternheimer linear response formalism~\cite{fabian,andrade07}. Thus, we can combine the potential response of unusual semi-local functionals with a plain LDA ground-state calculation to see just the effects of the xc response, and we can analyze and visualize potential responses and densities in real space to obtain a clear understanding of the functionals' properties.

\section{Functionals studied in this work}
\label{sec:tested_functionals}

The main interest of this work is the investigation of the linear response properties of AK13. 
However, in order to put the results into perspective, we also take a look at two well established, long-known GGAs: PBE as a paradigm example of a well-tested, usual GGA and the B88 GGA of Becke~\cite{b88}. The latter is of particular interest for our study because it has recently been pointed out that it shares several unusual features with AK13, such as divergences of the potential on nodal planes of the highest occupied orbital~\cite{thilo1,thilo2}. For the sake of completeness, we briefly summarize relevant aspects of these functionals in the following.

Exchange functionals of the GGA type are typically written in the form~\cite{Perdew_Wang_1986}
\begin{equation}
E_{\mathrm{x}}^{\text{\tiny SL}}[n] = A_{\mathrm{x}} \int\mathrm{d}^3 r \,\, n(\vect{r})^{\frac{4}{3}} F(s)
\label{equ:GGA_form_energy}
\end{equation}
where $F(s)$ is the exchange enhancement factor, \\$A_{\mathrm{x}} = - \frac{3}{4} \left(\frac{3}{\pi}\right)^{\frac{1}{3}} e^2$ and 
\begin{equation}
s = \frac{ \left| \nabla n(\vect{r}) \right| }{ 2 (3 \pi^2)^{\frac{1}{3}} n(\vect{r})^{\frac{4}{3}} }
\end{equation}
is a dimensionless density gradient. Different GGAs differ by different choices that are made for the enhancement factor.

The PBE functional's construction was guided by the aim to fulfill energetically relevant exact constraints, such as the homogeneous electron gas limit, proper coordinate scaling and the Lieb-Oxford bound. These constraints go along with an enhancement factor that goes to a constant for $s \rightarrow \infty$. A property of PBE that makes it a natural functional to compare AK13 to in the linear response context is the fact that PBE's enhancement factor was designed such that the functional recovers the linear response properties of LDA for the homogeneous electron gas. Therefore, the PBE response can be expected to be qualitatively similar to LDA in many cases. In other words, PBE is a GGA from which one expects predictable, unsurprising linear response properties. 

The B88 GGA is also considered a standard functional and it is a part of one of the most widely used hybrid functionals~\cite{b3lyp}. However, the guidelines along which B88 was designed are quite different from the PBE ones. The B88 functional was constructed such that it captures both the exact asymptotic behavior of the exchange energy density and the lowest-order gradient correction to LDA for small density gradients~\cite{b88}. In order to achieve this, the enhancement factor of B88 diverges for $s \rightarrow \infty$, yet in a way that has been called ``subcritical''~\cite{thilo1}, because despite the divergence of $F(s)$, the functional derivative of $E_\mathrm{x}^\mathrm{B88}[n]$ with respect to $n$ does not diverge for large distances from a finite's system center.

In contrast to the model potentials by which it was inspired, the AK13 approximation~\cite{ak13} is also based on the general GGA form of eq.~(\ref{equ:GGA_form_energy}). However, the guiding principles in its construction have not been energetic considerations~\cite{ak13rickard}. Instead, the aim in the design of AK13 was to make its functional derivative, i.e., the AK13 exchange potential, close to the Becke-Johnson model potential~\cite{bj}, which itself is in many respects a good model for the exchange-only Optimized Effective Potential. The most important property of the Becke-Johnson model which AK13 reproduces, is that asymptotically its potential for a finite system goes to a value that is determined by the highest occupied eigenvalue. In the AK13 functional, this is achieved by choosing $F(s)$ such that it diverges in a specific, ``critical'' manner~\cite{ak13}.
By letting the potential go to a finite, system-dependent value, step-structures are built into the potential which resemble the step-structures in the exchange-only Optimized Effective Potential that are related to the derivative discontinuity~\cite{mundtprl,leinkum}. As the derivative discontinuity is important for charge-transfer excitations~\cite{tozer03}, one may hope that a potential with such features may lead to a proper description of those.

\section{Linear response TDDFT in the Sternheimer approach}
\label{sec:Linear_response_TDDFT_calculations}

The most commonly used form of linear response TDDFT, often going by the name ``Casida formalism''~\cite{casida96,tddftbook2006p10}, is based on expanding the density response into particle-hole excitations. Here, we take a different route and solve the Sternheimer equations~\cite{tddftbook2006p10,andrade07}. So far, the Sternheimer approach is not as widely used as Casida's formalism, but it has the advantages of very efficient parallelizing~\cite{fabian} and of not requiring the explicit calculation of unoccupied orbitals.
As some of us have recently elsewhere presented the time-dependent Sternheimer approach in detail in the form that we also use here~\cite{fabian}, we can restrict ourselves to presenting the basic equations in the following.

In practice, a Sternheimer linear response calculation boils down to self-consistently solving the set of equations
\begin{equation}
\lbrack h_{\text{KS},\sigma}-\epsilon_{j \sigma}-\hbar\bar{\omega}\rbrack\phi_{j \sigma}^{+}
    =-\hat{Q}_j^{\sigma}\lbrack V_{\mathrm{ext}}^{+} + V_{\mathrm{Hxc}}^{+,\sigma}\rbrack\varphi_{j \sigma}
 \label{eq:SternPhi+}
\end{equation}
\begin{equation}
\lbrack h_{\text{KS},\sigma}-\epsilon_{j \sigma}+\hbar\bar{\omega}\rbrack\phi_{j \sigma}^{-}
   =-\hat{Q}_j^{\sigma}\lbrack V_{\mathrm{ext}}^{+} + V_{\mathrm{Hxc}}^{+,\sigma}\rbrack\varphi_{j \sigma}
 \label{eq:SternPhi-}
\end{equation}
for the orbital response components $\phi_{j \sigma}^{+}$, $\phi_{j \sigma}^{-}$ 
and excitation energies $\hbar \bar{\omega}$.
Here, 
\begin{equation}
h_{\mathrm{KS},\sigma} = -\frac{\hbar^2}{2m}\nabla^2 + \pot{ext}(\re) + \pot{Hxc}^{\sigma}(\re)
\label{eq:hKS}
\end{equation}
is the usual unperturbed ground-state Kohn-Sham Hamiltonian, with the Hartree and exchange-correlation (xc) contributions $\pot{Hxc}^{\sigma}(\vect{r}) = \pot{H}(\vect{r}) + \potxc^{\sigma}(\vect{r})$. The Kohn-Sham ground-state orbitals, which have been chosen to be real-valued, and eigenvalues of spin $\sigma$ are denoted by $\varphi_{j\sigma}(\re), \epsilon_{j\sigma}$, respectively. A finite $\eta \ll \omega $ is added~\cite{andrade07,fabian} to the excitation frequencies, i.e., $\bar{\omega} := \omega + i \eta$. This $\eta$ results from the adiabatic switch-on process, see below and Ref.~\cite{fabian}. It also improves the numerical stability of eqs.~(\ref{eq:SternPhi+}) and~(\ref{eq:SternPhi-}). $\hat{Q}_j^{\sigma}$ denotes the spin-dependent projector 
\begin{equation}
\hat{Q}_j^{\sigma} := 1 - \left| \varphi_{i \sigma} \rangle \langle \varphi_{i \sigma} \right| .
\label{eq:projector}
\end{equation}

The Fourier components $V_{\mathrm{ext}}^{+}$ of the external potential that appear on the right-hand sides of eqs.~(\ref{eq:SternPhi+}) and~(\ref{eq:SternPhi-}) are defined by the time-dependent, adiabatically applied, quasi-monochromatic external perturbation
\begin{equation}
\pot{ext}(\vect{r},t) = e^{\eta t} \left[ V_{\mathrm{ext}}^{+}(\vect{r}) e^{-i \omega t} + V_{\mathrm{ext}}^{-}(\vect{r}) e^{i \omega t} \right] ,
\label{eq:tdvext}
\end{equation}
where 
\begin{equation}
V_{\mathrm{ext}}^{+} = \left( V_{\mathrm{ext}}^{-} \right)^{*} .
\label{eq:vextcc}
\end{equation}
The Hartree- and xc-contributions 
\begin{equation}
V_{\mathrm{Hxc}}^{+,\sigma}=V_{\mathrm{H}}^{+}+V_{\mathrm{xc}}^{+,\sigma}
\label{eq:sumV+}
\end{equation}
appearing on the right-hand sides of eqs.~(\ref{eq:SternPhi+}) and~(\ref{eq:SternPhi-}) are obtained by solving the Poisson-like equation
\begin{equation}
\nabla^2V_{\mathrm{H}}^{+}= -4 \pi e^2 \left( n_\uparrow^{+} + n_\downarrow^{+} \right) 
 \label{eq:SternVH+}
\end{equation}
and computing
\begin{equation} 
V_{\mathrm{xc}}^{+,\sigma}(\vecvar{r})=\sum_{\tau = \uparrow,\downarrow} \int n_\tau^{+}(\vecvar{r'}) \fxc^{\sigma,\tau}(\vecvar{r},\vecvar{r'},\bar{\omega}) \, \mathrm{d}^3 r' .
 \label{eq:SternVxc+}
\end{equation}
Here,
\begin{equation}
\f{xc}^{\sigma,\tau}(\vect{r},\vect{r'},\bar{\omega}) := \int \f{xc}^{\sigma,\tau}(\vect{r},\vect{r'},t - t') e^{i \, \bar{\omega} (t - t')} \, \mathrm{d} (t - t')
\label{equ:Kernel_fourier_transform}
\end{equation}
is the Fourier transform of the exchange-correlation kernel
\begin{equation}
\f{xc}^{\sigma,\tau}(\vect{r},\vect{r'},t - t') := \funcder{\pot{xc}^{\sigma}[n_{\uparrow},n_{\downarrow}](\vect{r},t)]}{n_{\tau}(\vect{r'},t')}\Bigg\vert_{n_{\uparrow},n_{\downarrow}}.
\label{eq:defkernel}
\end{equation}
The density response
\begin{equation}
n_\sigma^{+}=\sum_{j=1}^{N_\sigma}\varphi_{j \sigma} \left(\phi_{j \sigma}^{+}+\phi_{j \sigma}^{-}\right)
\label{eq:Sternn+}
\end{equation}
enters into eqs.~(\ref{eq:SternVH+}) and~(\ref{eq:SternVxc+}), and thus a closed self-consistent cycle is obtained. 

The chosen xc approximation enters the Sternheimer equations in two places. First, it is part of the ground-state Hamiltonian of eq.~(\ref{eq:hKS}) and contributes to the eigenvalues and ground-state orbitals that feature in eqs.~(\ref{eq:SternPhi+}) and~(\ref{eq:SternPhi-}). Second, it determines the xc potential response of eq.~(\ref{eq:SternVxc+}). Using the linear response formalism instead of, e.g., a real-time propagation scheme~\cite{octopus2015,Lopata2011,Provorse2016,ingo2018} is decisive for studying the AK13 approximation's TDDFT performance, because AK13 ground-state orbitals are difficult to compute for finite, three-dimensional systems due to the previously discussed~\cite{thilo1,thilo2} particular features of the AK13 potential. However, in the linear response approach one can combine the ground-state orbitals of one exchange (and correlation) approximation with the kernel of some other approximation. In this way, we can test the kernel resulting from AK13.

In practice, we solve the Sternheimer equations by starting with eqs.~(\ref{eq:SternPhi+}) and~(\ref{eq:SternPhi-}) with merely $V_{\mathrm{ext}}^{+}$ on the right-hand side in order to generate initial values for the orbital responses $\phi_{j \sigma}^{+}$ and $\phi_{j \sigma}^{-}$, where our external perturbation is
\begin{equation}
V_{\mathrm{ext}}^{+}(\vect{r}) = e \left( \vect{E} \cdot \vect{r} \right) 
\label{eq:Def_Vext+}
\end{equation}
and $\vect{E}$ is a spatially homogeneous electric field. With $n_\sigma^{+}$ calculated according to eq.~(\ref{eq:Sternn+}) we obtain a complete 
set of quantities to start the self-consistency iteration by evaluating $V_{\mathrm{Hxc}}^{+,\sigma}$ via \mbox{eqs.~(\ref{eq:sumV+})-(\ref{eq:SternVxc+})}. Thus, 
we can construct the right-hand sides of eqs.~(\ref{eq:SternPhi+}) and~(\ref{eq:SternPhi-}) via eqs.~(\ref{eq:projector})-(\ref{eq:SternVxc+}) and from there calculate new versions of $\phi_{j \sigma}^{+}$ and $\phi_{j \sigma}^{-}$ by solving eqs.~(\ref{eq:SternPhi+}) and~(\ref{eq:SternPhi-}) again. These steps are iterated until a self-consistent solution is found. 

\section{Numerical results for Na$_2$}
\label{sec:Numerical_results_of_self_consistent_Na2_calculations}

Clusters of nearly-free-electron metals are in general reasonably well described by semi-local functionals~\cite{reinhard1999,Na5_03,Hofmann_Kuemmel_JCP_2012}. This is particularly true for sodium-clusters, as sodium (Na) is ``the nearly-free-electron metal par excellence"~\cite{Kuemmel_Andrae_Reinhard_2001}. For this reason, Na-clusters have often served as test systems for density functionals~\cite{rubioPRL96,marques2001,vasiliev1999,Vasiliev_Oeguet_Chelikowsky_2002}, 
and one can argue that a semi-local approximation should at least work for those.
If it passes this test, then further tests on more complicated systems are worthwhile, whereas testing it for more complicated systems makes little sense if already the simplest test, Na clusters, fails. In this logic, we here chose the dimer Na$_2$, which is known to be reasonably well described by (TD)LDA~\cite{vasiliev1999,Vasiliev_Oeguet_Chelikowsky_2002,kummelpra2001}, as the primary test system for which we evaluate the AK13 response.

The ground-state calculations were done with the Bay\-reuth version~\cite{mundttdpes} of the Parsec program~\cite{parsec}. The TDDFT calculations are based on a recently developed Sternheimer program package~\cite{fabian}. We used a Cartesian grid with a spacing of 0.45 Bohr ($a_0$) and sphere radii between 20 and 25 $a_0$, as indicated in the figure captions. The two Na atoms are located at \mbox{$x_1 = -2.9~a_0$} and \mbox{$x_2 = + 2.9~a_0$} on the \mbox{$x$-axis}
and are described by a Troullier-Martin~\cite{TM91} pseudopotential ($r_c=3.09 \, a_0$ for s-, p-, and d-shell). These parameters were chosen to ensure that the occupied as well as the first unoccupied eigenvalues of the ground-state calculation were converged to at least 10$^{-4}$ Rydberg (in the following, frequencies and potential responses are given in Rydberg atomic units),
and that the obtained TDLDA spectrum is in agreement with the one of Ref.~\cite{Vasiliev_Oeguet_Chelikowsky_2002}. The terms ``density response'', and ``potential response'' in the following refer to the \mbox{``$+$" Fourier components} unless stated otherwise.

For the reasons that have been discussed in detail in Ref.~\cite{thilo2}, using the AK13 GGA in self-consistent ground-state calculations is cumbersome and our attempts at converging such calculations have not been successful. Therefore, our interest here is not in using AK13 to set up the left-hand side of eq.~(\ref{eq:SternPhi+}), but in using AK13 for computing the potential response of eq.~(\ref{eq:SternVxc+}). In this way, by combining the AK13 response with a ``usual'' approximation for $h_\mathrm{KS}$, we can bring out the effects of the AK13 x kernel most clearly. For maximum transparency we chose the LDA for the ground-state Hamiltonian, with which we combine the AK13 x potential response. In order to calculate the latter, the AK13 kernel and potential response, respectively, have to be constructed in advance. These are calculated in section~\ref{appsec:fxc_of_tested_functionals} of the appendix.

However, we found that we could not converge the self-consistency iteration of the Sternheimer equations with the AK13 x potential response. 
\begin{figure}[t]
\includegraphics{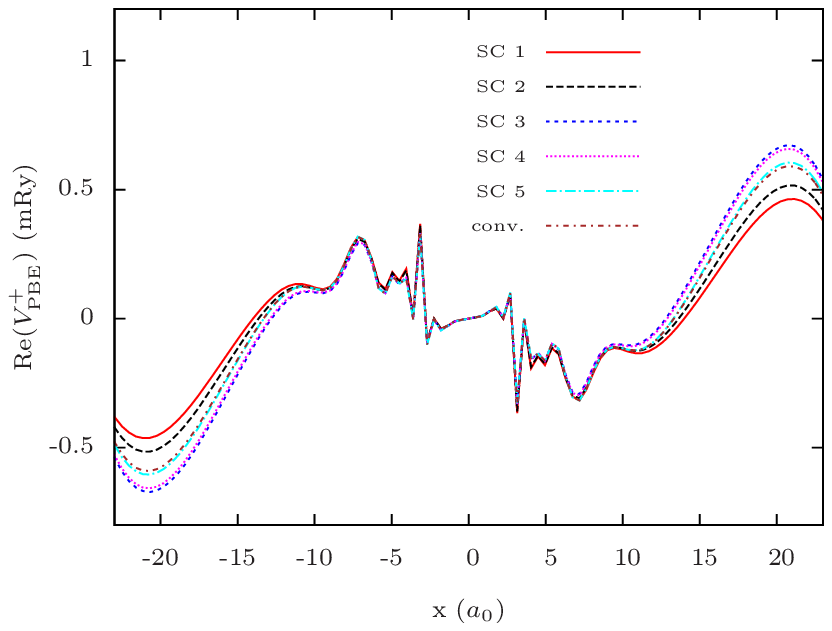}
\includegraphics{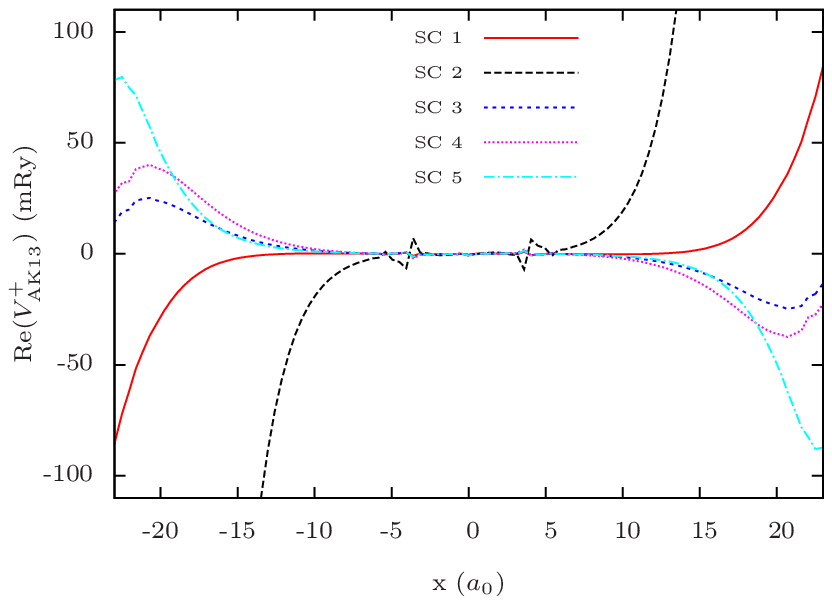}
\caption{
The real part of the x(c) response according to eq.~(\ref{eq:SternVxc+}) for
PBE (upper panel) and AK13 (lower panel) for an external electric field with polarization along the Cartesian (1,1,1) direction and an energy \mbox{$\hbar\omega = 0.3 \, \mathrm{Ry}$} for the first five self-consistency steps (SC). For PBE, also the converged result is shown. A boundary sphere with radius \mbox{$r = 25 \, a_0$} was used.
}
\label{fig:Comparison_convergence_beaviour_PBE_and_AK13}
\end{figure}
The lower panel of fig.~\ref{fig:Comparison_convergence_beaviour_PBE_and_AK13} shows the AK13 exchange potential response during the first five iterations of the Sternheimer equations. In the figures we omitted the last few grid points that lie close to the numerical boundary and are therefore affected by inaccuracies from the real-space finite differences.
It is evident that the changes of the AK13 potential response are enormous from one step to the next and the potential response even changes its sign. Oscillations build up at the boundaries of the simulation sphere and travel to the inside during the self-consistency iteration, impeding convergence. 
We tried to stabilize the numerical calculations in different ways, e.g., 
by starting the AK13 Sternheimer self-consistency iteration from a converged self-consistent LDA linear response calculation or using different mixing schemes.
However, none of the employed approaches nor combinations of them lead to a self-consistent, converging AK13 linear response calculation, even after several hundred iterations.

As a demonstration of how the xc response for a ``usual'' GGA looks like, 
the upper panel of fig.~\ref{fig:Comparison_convergence_beaviour_PBE_and_AK13} depicts the xc response of PBE. The PBE potential response differs relatively little from one self-consistency step to the next, and the Sternheimer iteration converges within nine steps. Thus, there is no problem with the GGA form in the Sternheimer approach per se, but something peculiar 
is happening in the AK13 calculation.

\begin{figure}[t]
\includegraphics{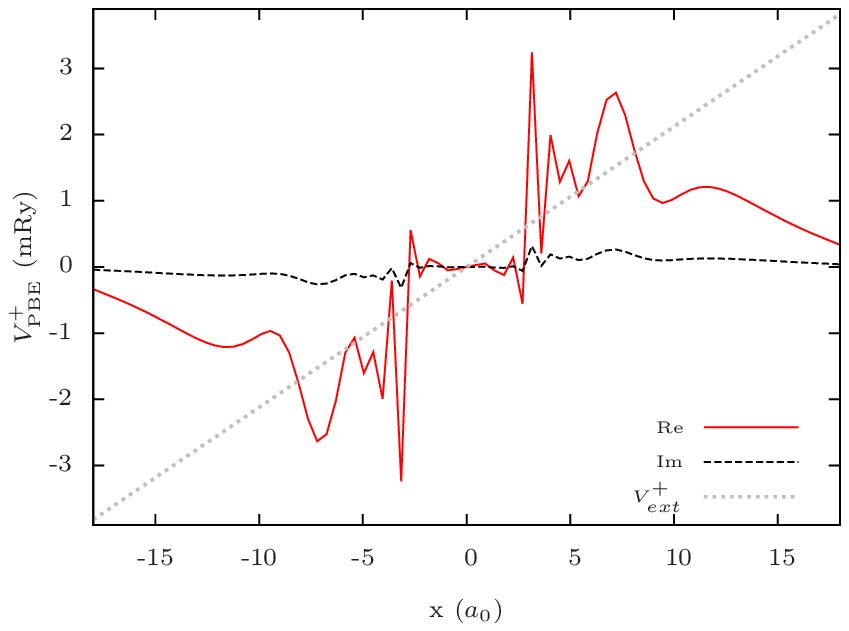}
\includegraphics{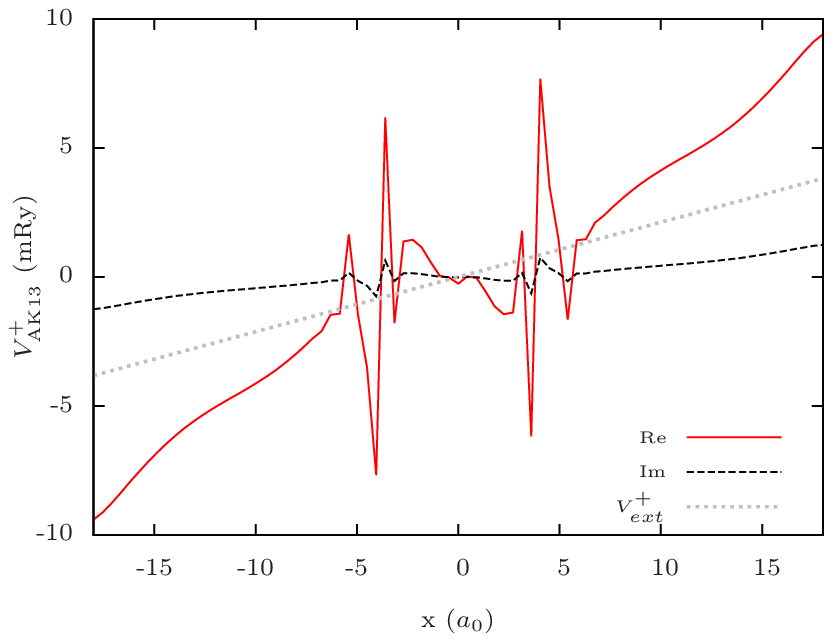}
\caption{Real and imaginary part of the potential response of PBE (upper panel) and AK13 (lower panel) along the \mbox{x-axis} for an external electric field with a polarization direction of (1,0,0) and energy \mbox{$\hbar\omega = 0.11 \, \mathrm{Ry}$}. 
The data is obtained by performing a self-consistent LDA ground-state and linear response calculation and subsequently evaluating the potential response for PBE and AK13, respectively, using the density and density response from the self-consistent LDA calculation. In addition, the potential of the applied external electric field is shown as a reference. A simulation sphere with radius \mbox{$r = 20 \, a_0$} was used.}
\label{fig:PBE_and_AK13_from_LDA_dir_100_Na2}
\end{figure}
In order to understand what is going on, we evaluated the potential response again in a different way. Instead of trying to analyze the self-consistent AK13 potential response, we performed a self-consistent LDA ground-state and linear response calculation and subsequently evaluated eq.~(\ref{eq:SternVxc+}) with the response-density obtained from LDA (which does not show any spurious features) and the x- and xc-kernel of AK13 and PBE, respectively. A striking feature of the AK13 response is revealed in this way. When the external electric field is applied in the (1,1,1) direction, the AK13 potential response exhibits an overall slope and a rising behavior towards the boundaries of the simulation sphere along all three coordinate axes. When changing the polarization direction to (1,0,0), the rising feature of the AK13 potential response vanishes along the y- and z-direction, but remains visible along the \mbox{x-axis}. The PBE response, on the contrary, always falls off to zero. Fig.~\ref{fig:PBE_and_AK13_from_LDA_dir_100_Na2} illustrates these findings, and also displays the potential of the external electric field as a reference. We stress that the observed features are numerically stable and not artifacts of how the potentials and densities are numerically computed.

Summarizing these findings we note that the direction of the AK13's potential response slope depends on the direction of the external electrical field, and the slope is proportional to its modulus. 
The real part of the AK13 potential response becomes larger than the potential of the external electric field for large distances, making it the asymptotically leading term. These somewhat surprising findings call for further explanation. To this end, we take a look at the hydrogen atom, for which exact relations for the exchange response can be derived as shown below.

\section{Analytical analysis of the exchange potential response}

In the following section we contrast the exact 
analytical result for the hydrogen atom response with the one obtained from the different functionals.

\subsection{Asymptotics of the exact exchange potential response}
\label{subsec:Analytical_asymptotics_of_the_potential_response}

One may argue that a one-electron system is quite a challenging test for a semi-local functional because of the well-known self-interaction problem, i.e., one might argue that failing the one-electron test may not necessarily imply that a semi-local approximation is useless. E.g., the LDA ground-state energy for the hydrogen (H) atom is not particularly accurate, yet LDA is nonetheless a useful approximation for a lot of 
many-electron systems. However, for our present purposes the H-atom is a good test case, and a very relevant one, because our aim here is not testing quantitative performance, but understanding qualitative features of the exchange response. For this, the H-atom is ideal because the exact potential response can easily be derived.

For every one-electron system the exact exchange functional just cancels the Hartree contribution~\cite{Primer_in_DFT}. Thus, in this case the exact exchange potential is the negative Hartree potential,
\begin{equation}
\pot{x}^{\mathrm{ex}}(\vect{r},t) = - \, \pot{H}(\vect{r},t) = - e^2 \int\mathrm{d}^3 r' \, \frac{n(\vect{r'},t)}{\left| \vect{r} - \vect{r'} \right|}, 
\label{equ:v_x_ex_one_electron_system}
\end{equation} 
and consequently the exact x kernel is also just the negative Hartree kernel, from which the potential response
\begin{equation}
V_{\mathrm{x, ex}}^{+}(\vect{r}) = - V_{\mathrm{H}}^{+}(\vect{r}) = - e^2 \int\mathrm{d}^3 r' \, \frac{n^{+}(\vect{r'})}{\left| \vect{r} - \vect{r'} \right|}
\label{equ:V_x_ex_+_one_electron_system}
\end{equation}
follows.
From eq.~(\ref{equ:V_x_ex_+_one_electron_system}) the asymptotic behavior of the exact x potential response can be determined via a multipole-expansion
\begin{align}
\begin{split}
V_{\mathrm{x, ex}}^{+}(\vect{r}) = & - e^2 \int\mathrm{d}^3 r' \, \frac{n^{+}(\vect{r'})}{\left| \vect{r} - \vect{r'} \right|} = - e^2 \frac{\overbrace{\int n^{+}(\vect{r'}) \, \mathrm{d}^3 r'}^{=0}}{r} \\
& + e \frac{\overbrace{- e \int \vect{r'} \, n^{+}(\vect{r'}) \mathrm{d}^3 r'}^{=:\vect{p}^{+}} \, \cdot \, \vect{r}}{r^3} + \mathcal{O}\left(\frac{1}{r^3}\right) \\[1ex]
= & e \frac{\vect{p}^{+} \cdot \vect{e_r}}{r^2} + \mathcal{O}\left(\frac{1}{r^3}\right) \, \xrightarrow{r \rightarrow \infty} \, 0 \, ,
\end{split}
\label{equ:V_x_ex_+_one_electron_system_multipole_expansion}
\end{align}
where the density response integrates to zero due to particle number conservation and $\vect{p}^{+} := - e \int \vect{r'} \, n^{+}(\vect{r'}) \mathrm{d}^3 r'$ is the dipole moment of the density response (i.e., the transition dipole).
Thus, the exact exchange potential response tends to zero asymptotically proportional to $\frac{1}{r^2}$ or faster. In directions perpendicular to the dipole moment $\vect{p}^{+}$ it decays proportionally to at least $\frac{1}{r^3}$.

\subsection{Asymptotics of the potential response of PBE, AK13, and B88}
\label{sec:Asymptotics_of_the_potential_response_of_semi_local_exchange_approximations}

In order to calculate the asymptotic behavior for the exchange-correlation approximations that we want to compare to eq.~(\ref{equ:V_x_ex_+_one_electron_system_multipole_expansion}), the asymptotics of the density response is required, cf.\ eq.~(\ref{eq:SternVxc+}). For the transition from the 1s orbital to the 2p$_x$ orbital of a hydrogen atom, it is given by
\begin{equation}
n^{+}(\vect{r}) = \varphi_{1 \mathrm{s}}(\vect{r}) \, \varphi_{2 {\mathrm{p}_x}}(\vect{r}),
\end{equation}
or explicitly,
\begin{equation}
n(\vect{r})  = \frac{1}{a_0^3 \pi} e^{- 2 \frac{r}{a_0}} 
\label{equ:One_electron_density_explicit_form}
\end{equation}
and
\begin{equation}
n^{+}(\vect{r})  = \frac{1}{a_0^3 \pi \sqrt{32}} \frac{x}{a_0} e^{- \frac{3}{2} \frac{r}{a_0}}.
\label{equ:One_electron_density_response_explicit_form}
\end{equation}
In appendix~\ref{app:1s2pstern} we derive this relation from the Sternheimer equations.

Based on this density response we can proceed to evaluate the potential response of the PBE, AK13 and B88 approximations. Since all of these originate from the GGA form~(\ref{equ:GGA_form_energy}),
we calculate the elements required for the evaluation of the potential response asymptotics for the general GGA form in appendix~\ref{appsec:fxc_of_tested_functionals}. One then just has to insert the enhancement factors $F(s)$ for PBE , 
AK13 and B88, respectively, into the resulting equations to obtain the potential response for these functionals. 

The important equations are eqs.~(\ref{eq:Potential_response_SL_final_form}) and~(\ref{eq:I-II-III_final_form}). Together with eqs.~(\ref{eq:s+})~-~(\ref{eq:t+}) they allow expressing the potential response $V_{\mathrm{x},\text{\tiny SL}}^{+}(\vect{r})$, where SL stands for PBE, AK13
and B88, in terms of the density, the density response and derivatives of these two.
According to eqs.~(\ref{equ:One_electron_density_explicit_form}) and~(\ref{equ:One_electron_density_response_explicit_form}), in the H-atom calculation 
the density is spherically symmetric, but the density response only exhibits cylindrical symmetry around the \mbox{$x$-axis}. Therefore, we calculate all derivatives in cylindrical coordinates. The gradient and Laplacian of the density for the H-atom case are (cf.\ eq.~(\ref{equ:One_electron_density_explicit_form}))
\begin{align}
\nabla n(\vect{r}) = \, & \frac{2}{a_0} \left( \vect{e}_{\rho} \frac{- \rho}{r} + \vect{e}_x \frac{-x}{r} \right) n(\vect{r}) \notag\\[2ex]
 =: & - \frac{2}{a_0} \, \vect{e}_{r} \, n(\vect{r}) 
\label{equ:Grad_n_one_electron} \\[2ex]
\nabla^2 n(&\vect{r}) = \frac{4}{a_0^2} \left( 1 - \frac{a_0}{r} \right) n(\vect{r}) ,
\label{equ:Lapl_n_one_electron}
\end{align}
where \mbox{$r^2 = x^2 + y^2 + z^2 = x^2 + \rho^2$}, $\vect{e}_{\rho}$ 
and $\vect{e}_{x}$ are the corresponding unit vectors in cylindrical coordinates and \\$\vect{e}_{r} := \vect{e}_{\rho} \frac{\rho}{r} + \vect{e}_x \frac{x}{r}$.\\

With these two equations, we calculate the reduced density gradients (cf.\ eqs.~(\ref{eq:Definition_s}),~(\ref{eq:Definition_u}) and~(\ref{eq:Definition_t}))
\begin{align}
s = & \, \frac{2}{a_0} \frac{1}{2 \left(3 \pi^2 \right)^{\frac{1}{3}}} \, n(\vect{r})^{-\frac{1}{3}} 
\label{equ:s_one_electron}\\[2ex]
u = & \, \frac{8}{a_0^3} \frac{1}{8 \left(3 \pi^2\right)} \, n(\vect{r})^{-1} = s^3
\label{equ:u_one_electron}\\[2ex]
t = & \, \frac{4}{a_0^2} \,\, \frac{\,\, 1 - \frac{a_0}{r}\,\,}{4 \left(3 \pi^2\right)^{\frac{2}{3}}} \, n(\vect{r})^{-\frac{2}{3}} \notag\\
 = & \, s^2 \left( 1 - \frac{a_0}{r} \right)
\label{equ:t_one_electron}
\end{align}
The derivatives of the density response are (cf.\ eq.~(\ref{equ:One_electron_density_response_explicit_form}))
\begin{align}
\nabla & n^{+}(\vect{r}) = \frac{1}{a_0} \left( - \frac{3}{2} \, \vect{e}_{r} + \frac{a_0}{x} \, \vect{e}_{x} \right) \, n^{+}(\vect{r})
\label{equ:Grad_n+_one_electron} \\[2ex]
\nabla^2 & n^{+}(\vect{r}) = \frac{1}{a_0^2} \left( \frac{9}{4} - 6 \frac{a_0}{r} \right) \, n^{+}(\vect{r}), 
\label{equ:Lapl_n+_one_electron}
\end{align}
and analogously we obtain (cf.\ eqs.~(\ref{eq:s+}),~(\ref{eq:u+}) and~(\ref{eq:t+})) 
\begin{align}
s^{+}(\vect{r}) = & \,\, - \frac{1}{a_0} \frac{\frac{7}{6} + \frac{a_0}{r}}{2 \left(3 \pi^2 \right)^{\frac{1}{3}}} \quad\;\, \frac{n^{+}(\vect{r})}{n(\vect{r})^{\frac{4}{3}}}
\label{equ:s+_one_electron}\\[3ex]
u^{+}(\vect{r}) = & \,\, - \frac{1}{a_0^3} \frac{\frac{27}{2} + 10 \frac{a_0}{r}}{8 \left(3 \pi^2\right)} \quad \frac{n^{+}(\vect{r})}{n(\vect{r})^2} 
\label{equ:u+_one_electron}\\[3ex]
t^{+}(\vect{r}) = & \,\, - \frac{1}{a_0^2} \frac{\frac{53}{12} - \frac{2}{3} \frac{a_0}{r}}{4 \left(3 \pi^2\right)^{\frac{2}{3}}} \;\, \frac{n^{+}(\vect{r})}{n(\vect{r})^{\frac{5}{3}}}
\label{equ:t+_one_electron}
\end{align}
Thus, we have derived all quantities that allow to evaluate the general form of the GGA potential response from 
eqs.~(\ref{eq:Potential_response_SL_final_form}) and~(\ref{eq:I-II-III_final_form}) for the $1 \mathrm{s} \rightarrow 2 \mathrm{p}_x$ excitation.
Inserting the appropriate enhancement factors we find the asymptotic behavior of the PBE potential response as
\begin{equation}
V_{\mathrm{x},\text{\tiny PBE}}^{+}(\vect{r}) \xrightarrow{r \rightarrow \infty} \left( 1 + \kappa \right) \, V_{\mathrm{x},\text{\tiny LDA}}^{+}(\vect{r}) , 
\label{equ:Asymptotic_behavior_PBE_potential_response_one_electron}
\end{equation}
where $\kappa$ is the parameter fixed in the PBE construction~\cite{pbe} and
\begin{equation}
V_{\mathrm{x},\text{\tiny LDA}}^{+}(\vect{r}) \xrightarrow{r \rightarrow \infty} \frac{4}{9} \, A_{x} \frac{n^{+}(\vect{r})}{n(\vect{r})^{\frac{2}{3}}}.
\label{equ:Asymptotic_behavior_LDA_potential_response_one_electron}
\end{equation}
(Formally, LDA corresponds to the general GGA form of eq.~(\ref{equ:GGA_form_energy}) with $F_{\text{\tiny LDA}}(s) \equiv 1$.)
The corresponding asymptotical result (with $B_1$ being the parameter from the AK13 construction~\cite{ak13}) for the AK13 potential response is 
\begin{equation}
V_{\mathrm{x},\text{\tiny AK13}}^{+}(\vect{r}) \xrightarrow{r \rightarrow \infty} \frac{91}{144} \, A_{x} \, \frac{B_{1}}{a_0} \, \frac{1}{\left(3 \pi^2 \right)^{\frac{1}{3}}} \frac{n^{+}(\vect{r})}{n(\vect{r})}, 
\label{equ:Asymptotic_behavior_AK13_potential_response_one_electron}
\end{equation}
and for B88 we obtain
\begin{equation}
 V_{\mathrm{x},\text{\tiny B88}}^{+}(\vect{r}) \xrightarrow{r \rightarrow \infty} \frac{313}{96} \frac{a_0 e^2}{r^2} \frac{n^{+}(\vect{r})}{n(\vect{r})}
\label{equ:Asymptotic_behavior_B88_potential_response_one_electron}
\end{equation}

Comparing eqs.~(\ref{equ:Asymptotic_behavior_PBE_potential_response_one_electron}) -~(\ref{equ:Asymptotic_behavior_B88_potential_response_one_electron}) shows that the PBE response, as expected, is similar to the LDA one, but that the AK13 and B88 responses differ markedly. 
Inserting eqs.~(\ref{equ:One_electron_density_explicit_form}) and~(\ref{equ:One_electron_density_response_explicit_form}) we can determine the asymptotics of the potential response for the $1 \mathrm{s} \rightarrow 2 \mathrm{p}_x$ excitation. For PBE, it falls of to zero just like LDA, 
\begin{equation}
\lim\limits_{r \rightarrow \infty} V_{\mathrm{x},\text{\tiny PBE}}^{+}(\vect{r}) \propto x \, e^{- \frac{1}{6} \frac{r}{a_0}} \xrightarrow{r \rightarrow \infty} 0.
\label{equ:Asymptotic_behavior_PBE_potential_response_exponential_one_electron}
\end{equation}
whereas for AK13 we find
\begin{equation}
\lim\limits_{r \rightarrow \infty} V_{\mathrm{x},\text{\tiny AK13}}^{+}(\vect{r}) \propto \; x \, e^{\frac{1}{2} \frac{r}{a_0}} \xrightarrow{r \rightarrow \infty} \infty,
\label{equ:Asymptotic_behavior_AK13_potential_response_exponential_one_electron}
\end{equation}
and for B88
\begin{equation}
\lim\limits_{r \rightarrow \infty} V_{\mathrm{x},\text{\tiny B88}}^{+}(\vect{r}) \propto \frac{x}{r^2} \, e^{\frac{1}{2} \frac{r}{a_0}} \xrightarrow{r \rightarrow \infty} \infty.
\label{equ:Asymptotic_behavior_B88_potential_response_exponential_one_electron}
\end{equation}

Comparing this to the exact result given in eq.~(\ref{equ:V_x_ex_+_one_electron_system_multipole_expansion}), we see that for the studied excitation the PBE response, although falling off too rapidly, goes to the correct limiting value (zero). The one of AK13 and B88, however, grows exponentially.  
We thus find that neither B88 nor AK13 are modeling the exact exchange reponse well for this one electron transition. The strength of the divergence observed for AK13 is also an important step in understanding the numerical convergence problems.

\subsection{Numerical confirmation}
\label{subsec:Numerical confirmation}

\begin{figure}[t!]
\includegraphics{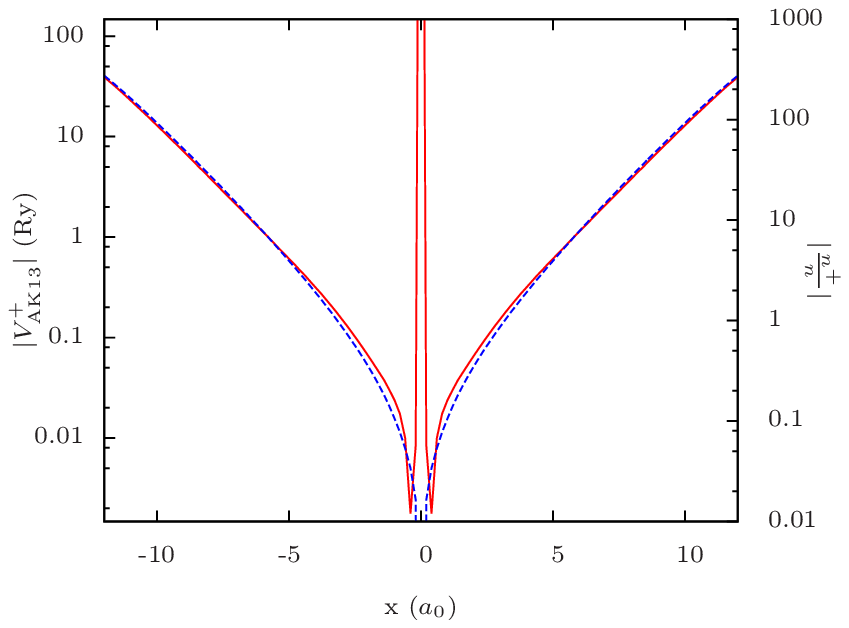}\vspace{1ex}
\includegraphics{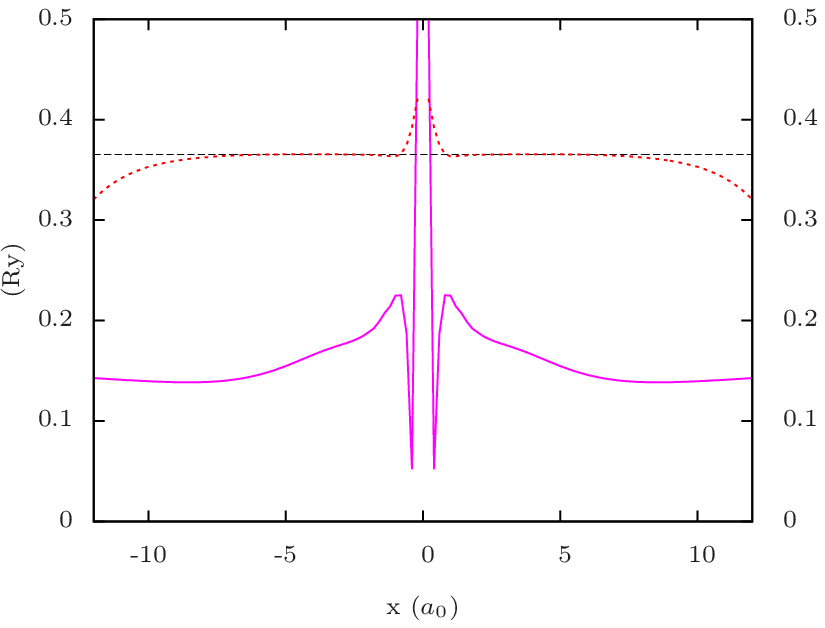}
\caption[]{Quasi-exact numerical ground-state and response calculation of the hydrogen atom. A sphere with radius $r = 15 \, a_0$ and a grid spacing of $\Delta r = 0.20 \, a_0$ was used.
Upper panel, red curve, plotted against left ordinate: Absolute value of the AK13 potential response $\vert V_{\mathrm{\tiny AK13}}^{+} \vert$ evaluated with the ground-state density and density response of the quasi-exact numerical calculation for the hydrogen atom. Dashed blue 
curve, plotted against right ordinate: Ratio of the density response to the density $\vert \frac{n^{+}}{n}\vert$. The plot shall demonstrate that both functions are proportional to each other.\\
Lower panel, 
solid magenta curve, plotted against left ordinate: Numerical data for $\vert V_{\mathrm{\tiny AK13}}^{+} \vert $ divided by the numercial data for $ \vert \frac{n^{+}}{n} \vert$. 
Dotted red curve, plotted against right ordinate: Numerical data for $\vert \frac{n^{+}}{n} \vert$ divided by the analytical data for $\vert \frac{n^{+}}{n} \vert$. The dashed black line serves as a reference for perfect proportionality between analytical and numerical data.}
\label{fig:AK13_one_electron_exact}
\end{figure}

Finally, in order to really rule out that our non-converging AK13 calculations in section~\ref{sec:Numerical_results_of_self_consistent_Na2_calculations} are a consequence of numerical issues in our Sternheimer implementation, we check our numerics by reproducing the just derived analytical result with our Sternheimer program. To this end, we do a numerical quasi-exact ground-state calculation of the hydrogen atom with the code used in section~\ref{sec:Numerical_results_of_self_consistent_Na2_calculations}, and also do the linear response calculation quasi-exactly for the hydrogen atom. By quasi-exact we mean that numerical convergence parameters were chosen very stringent and only the external perturbation potential in the Sternheimer eqs.~(\ref{eq:SternPhi+}) and~(\ref{eq:SternPhi-}) is taken into account, which is the exact situation for the hydrogen atom. The hydrogen atom was described using 
a Troullier-Martin~\cite{TM91} pseudopotential ($r_c=1.39 a_0$), 
and we tested that with this pseudopotential energies and eigenvalues are close to the ones from the true hydrogen potential. 
With the thus numerically obtained density response we numerically evaluate the AK13 potential response. The result is depicted in fig.~\ref{fig:AK13_one_electron_exact}.

According to eq.~(\ref{equ:Asymptotic_behavior_AK13_potential_response_one_electron})
the AK13 potential response is expected to be proportional to the ratio of the density response and 
the density. The upper panel of fig.~\ref{fig:AK13_one_electron_exact} shows the absolute value of these two quantities on a logarithmic scale. Over a wide region of space we find close agreement. In the interior and in the outer region of the displayed simulation volume the two curves slightly deviate from each other. The dotted red line in the lower panel of fig.~\ref{fig:AK13_one_electron_exact} shows that this is a consequence of the numerical and analytical results for $\frac{n^{+}(\vect{r})}{n(\vect{r})}$ deviating from each other in the center of the grid and close to the boundaries. These deviations are expected and easily understood. The deviation in the interior is expected because of the use of a pseudopotential and the finite discretization, which lead to a numerical ground-state density that lacks the exact cusp at the nuclear position (x=0), as in every pseudopotential calculation.
The deviations close to the grid boundary are a consequence of the necessity of enforcing the
zero-boundary condition in the calculation of the ground-state orbitals. As the analytical density vanishes asymptotically and thus never becomes zero exactly, the numerical data has to slightly deviate from the correct asymptotic behavior near the boundary.

However, the important observation in fig.~\ref{fig:AK13_one_electron_exact} is that the numerical evaluation of the AK13 response does show the same behavior as the analytical evaluation in all regions of space where it can be expected to show it (i.e., in those regions of space where the analytical and the numerical density are close to each other). The solid magenta curve in the lower panel of fig.~\ref{fig:AK13_one_electron_exact} confirms that the ratio $\vert V_{\mathrm{\tiny AK13}}^{+} \vert/ \vert \frac{n^{+}}{n} \vert$ tends to a constant for large values of $x$, as it should.
Therefore, we confirm the reliability of our Sternheimer implementation, and also confirm the conclusion that the non-converging Sternheimer iterations for AK13 are not a result of numerical problems, but are to be attributed to the strongly diverging response of the AK13 approximation.

\begin{figure}[t]
\includegraphics{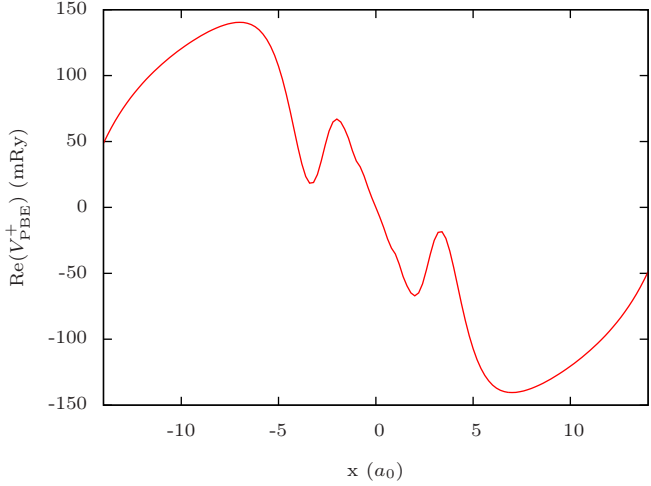}
\caption{PBE potential response evaluated with the ground-state density and density response of the quasi-exact numerical calculation for the hydrogen atom.
}
\label{fig:PBE_potential_response_H_atom}
\end{figure}

For the sake of completeness we depict the PBE potential response in fig.~\ref{fig:PBE_potential_response_H_atom}.
The figure is in line with the analytical result for PBE and shows that the PBE response does not show any divergences. 

\begin{figure}[t]
\includegraphics{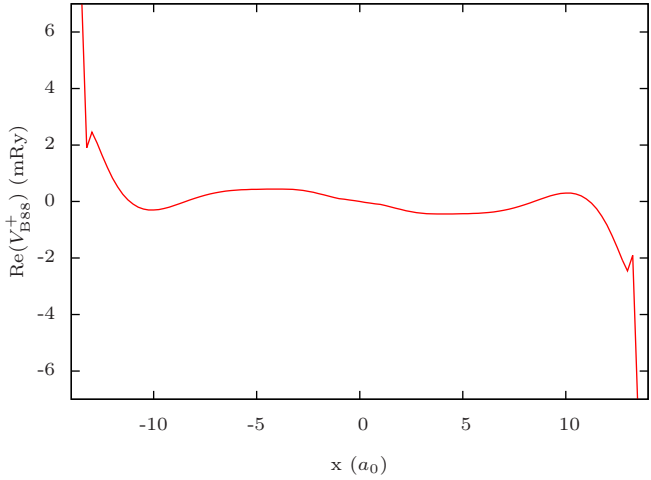}
\caption{B88 potential response evaluated with the ground-state density and density response of the quasi-exact numerical calculation for the hydrogen atom.
}
\label{fig:B88_potential_response_H_atom}
\end{figure}

Finally, we take a look at the B88 response. As explained earlier, one has to keep in mind that B88, like AK13, is built with a diverging enhancement factor, yet the divergence is milder.  Fig.~\ref{fig:B88_potential_response_H_atom} depicts the potential response for B88, evaluated in the same way on the numerical, quasi-exact density and density response as just described for AK13.
The potential response is smooth in regions of space where the density is high. Close to the grid boundary we observe a strong rise 
and see a spike that we attribute to the influence of the grid boundary on the finite differences.
However, these features do not hinder convergence of the Sternheimer equations with the B88 approximation. 
We could obtain fully self-consistent, converged Sternheimer results for the H-atom $1 \mathrm{s} \rightarrow 2 \mathrm{p}_x$ excitation for B88. The excitation energy is not too different to the one found with xLDA or xPBE. On a grid of radius $r = 15 \, a_0$ and with a grid spacing of $\Delta r = 0.20 \, a_0$, which leads to a numerical accuracy of a few mRy, we find excitation energies of 542, 572, 575 mRy for xLDA, xPBE and B88, respectively.

\begin{figure}[h]
\includegraphics{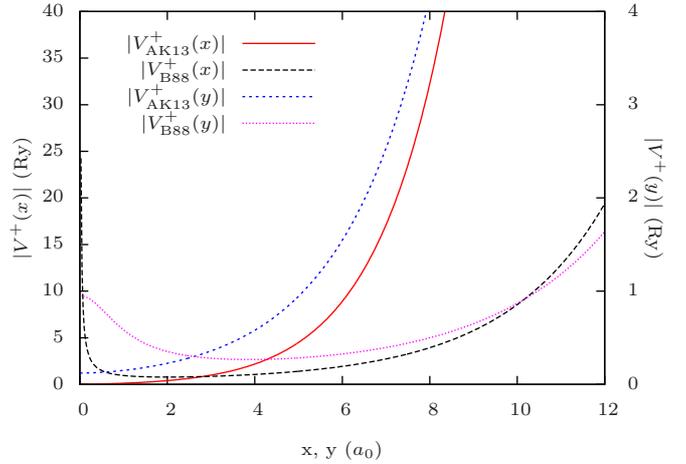}
\caption{Analytical asymptotic behavior of the AK13 and B88 functional according to eqs.~(\ref{equ:Asymptotic_behavior_AK13_potential_response_one_electron}) and~(\ref{equ:Asymptotic_behavior_B88_potential_response_one_electron}). The solid red and dashed black curves are plotted against the left ordinate and depict the data along the x-direction for \mbox{(y,z) = (0,0) $a_0$}. The dashed blue and dashed magenta curves are plotted against the right ordinate and show the data along the y-direction for \mbox{(x,z) = (1,0) $a_0$}.
The offset of 1 $a_0$ in the $x$-direction was chosen to avoid the orbital nodal plane.
}
\label{fig:Asymptotics_AK13_vs_B88}
\end{figure}
Thus, the B88 response calculations show that a diverging enhancement factor and potential response need not necessarily lead to problems in TDDFT calculations. In order to clarify the situation further, in fig.~\ref{fig:Asymptotics_AK13_vs_B88} we depict the asymptotics of both functionals, i.e., fig.~\ref{fig:Asymptotics_AK13_vs_B88} visualizes eqs.~(\ref{equ:Asymptotic_behavior_AK13_potential_response_one_electron}) and~(\ref{equ:Asymptotic_behavior_B88_potential_response_one_electron}) evaluated for the exact density and density response. The potential response of both functionals rises with the same exponential rate in the asymptotic limit, but the one of B88 is moderated by $\frac{1}{r^2}$. Fig.~\ref{fig:Asymptotics_AK13_vs_B88} shows that this leads to a considerably slower rise.
This finding is in line with earlier observations for ground-state calculations~\cite{thilo1}: Although both AK13 and B88 diverge on some orbital nodal planes, it is possible to converge ground-state calculations for B88 but not for AK13. The milder divergence of the B88 ground-state potential can be numerically covered, whereas the pronounced divergence of AK13 leads to serious problems. From our Sternheimer results we conclude that the situation is similar for the potential response.

\section{Conclusion}
\label{sec:conclusion}

We investigated the linear response of the AK13 GGA with the aim of exploring whether the unusual features of its functional derivative can be exploited beneficially in TDDFT calculations. We found that we cannot converge such calculations. Contrasting the AK13 response with the one of the PBE GGA, for which the Sternheimer equations can be solved without any
problem, revealed that AK13 leads to an asymptotically 
increasing exchange 
response that is absent in PBE. By comparing this
to the exact response, which we calculated for the hydrogen atom, we traced this finding back to AK13's diverging enhancement factor
and identified the feature as not being in agreement with the proper exchange behavior.
Comparison with the B88 exchange GGA, which also has a diverging enhancement factor but leads to a self-consistent solution of the Sternheimer equations, showed that a diverging enhancement factor in itself does not need to 
ruin the response properties, but the particular form that is chosen in AK13 is problematic for TDDFT applications.

Our original hope was that the AK13 functional may have been useful for providing ``kernel corrections'' to the linear response in situations where usual GGAs, which closely follow the density, do not. 
Long-range charge-transfer excitations would have been a hallmark example. Our study showed that even much simpler excitations cannot be calculated with the adiabatic AK13 functional. The peculiar results found here for AK13 indicate that it is very difficult to develop a semi-local functional that leads to pronounced but beneficial response properties. Whereas it is clear that the GGA potential response would have to be sharply increasing in regions of vanishing orbital overlap in order to provide a non-vanishing correction, our results here showed that too much of a divergence can ruin the response properties altogether. A possible way out of this disaccord may be to try to model the response semi-locally, but not semi-locally in the density, but semi-locally in the orbitals, such as done in meta-GGAs~\cite{Nazarov2011}. In this way, it may be possible to obtain finite ``kernel corrections'' in a different manner, namely not by providing a
potential with diverging properties, 
but by providing a relative potential offset of the donor- and acceptor regions of a charge-transfer system.

\section*{Acknowledgments}
S.K.\ and J.G.\ acknowledge support by Deutsche Forschungsgemeinschaft Graduiertenkolleg 1640 and by the study program ``Biological Physics`` of the Elite Network of Bavaria. S.K. further acknowledges support by the German-Israeli Foundation for Scientific Research and Development. J.G. and F.H. acknowledge support by the University of Bayreuth Graduate School.  R.A. acknowledges support from the Swedish Research Council (VR) project no. 2016-04810 and the Swedish e-Science Research Centre (SeRC).

S.K.\ thanks E.K.U.\ Gross for countless discussions at conferences and for infecting him with his enthusiasm for TDDFT.

\section*{Author contribution statement}
J.G. did the analytical and numerical calculations including the programming of the AK13 response routines, F.H. developed and programmed the Sternheimer linear response code, S.K. initiated and supervised the project, J.G. and S.K. analyzed the results and structured the discussion, all authors contributed to the discussion, the writing and the proofreading of the manuscript.

\section{Appendix}

\subsection{Details of how the GGA kernel enters the Sternheimer equations}
\label{appsec:fxc_of_tested_functionals}

We here give some details about how to use functionals of the GGA form in the Sternheimer approach.
It is an appealing feature of the Sternheimer equations that they do not require the xc kernel by itself, but only the xc potential response. This is advantageous as the potential response only depends on one three-dimensional spatial coordinate, whereas the xc kernel depends on two. 
Thus, it is easier to analyze the effects of a particular xc approximation on the linear response by looking at the xc potential response instead of the xc kernel itself. 

The starting point for the derivation of the AK13 potential response is the GGA form of the (semi-local) exchange energy functional from eq.~(\ref{equ:GGA_form_energy}).
The corresponding potential to this is the functional derivative $\funcder{E_{\mathrm{x}}^{\text{\tiny SL}}[n]}{n(\vect{r})}$, which can be extracted from Ref.~\cite{Perdew_Wang_1986} and rearranged to
\begin{eqnarray}
&& \pot{x}^{\text{\tiny SL}}[n](\vect{r}) = \frac{4}{3} \, A_{\mathrm{x}} \, n(\vect{r})^{\frac{1}{3}} \label{eq:GGA_form_potential}\\[1ex] 
&& \times \underbrace{ \left[ F(s) - \left( \frac{3}{4} \, \frac{t}{s} - \frac{3}{4} \, \frac{u}{s^2} + s \right) F'(s) - \left( \frac{3}{4} \, \frac{u}{s} - s^2 \right) F''(s) \right] }_{=: B(\vect{r})} \nonumber
\end{eqnarray}
with 
\begin{eqnarray}
s = && \,\, \frac{|\gradn|}{2 \left(3 \pi^2 \right)^{\frac{1}{3}} n(\vect{r})^{\frac{4}{3}}}
\label{eq:Definition_s}\\[2ex]
u = && \,\, \frac{\gradn \cdot \nabla |\gradn|}{8 \left(3 \pi^2\right) n(\vect{r})^3}
\label{eq:Definition_u}\\[2ex]
t = && \,\, \frac{\nabla^2 n(\vect{r})}{4 \left(3 \pi^2\right)^{\frac{2}{3}} n(\vect{r})^{\frac{5}{3}}}
\label{eq:Definition_t}
\end{eqnarray}
According to eq.~(\ref{eq:defkernel}) the kernel is the functional derivative of the potential. Thus, it takes the form
\begin{equation}
\begin{split}
\f{x}^{\text{\tiny SL}}[n](\vect{r},\vect{r'}) = & \frac{4}{9} \, A_{\text{x}} \, n(\vect{r})^{-\frac{2}{3}} \, \delta(\vect{r}-\vect{r'}) \, B(\vect{r}) \\[1ex] 
& + \frac{4}{3} \, A_{\text{x}} \, n(\vect{r})^{\frac{1}{3}} \, \funcder{B(\vect{r})}{n(\vect{r'})}
\end{split}
\label{eq:Kernel_SL_GGA}
\end{equation}
with
\begin{eqnarray}
B(\vect{r}) = F(s) && - \left( \frac{3}{4} \, \frac{t}{s} - \frac{3}{4} \, \frac{u}{s^2} + s \right) F'(s) \nonumber\\[1ex]
&& -  \left( \frac{3}{4} \, \frac{u}{s} - s^2 \right) F''(s) .
\label{eq:Definition_B}
\end{eqnarray}
Aside from derivatives of the exchange enhancement factor, the functional derivatives of $s$, $u$ and $t$ are needed for $\funcder{B(\vect{r})}{n(\vect{r'})}$. These are given by:
\begin{eqnarray}
\funcder{s(\vect{r})}{n(\vect{r'})} = && \,\, \frac{n(\vect{r}) \, \gradn \cdot \gradd - \frac{4}{3} \, \left| \gradn \right|^2 \, \delta(\vect{r}-\vect{r'})}{2 \left(3 \pi^2 \right)^{\frac{1}{3}} n(\vect{r})^{\frac{7}{3}} \, \left| \gradn \right|} \nonumber\\
\label{eq:Definition_ds_dn}\\[3ex]
\funcder{u(\vect{r})}{n(\vect{r'})} = && \,\, \frac{n(\vect{r}) \, \nabla |\gradn| \cdot \gradd}{8 \left(3 \pi^2\right) n(\vect{r})^4} \nonumber\\[2ex]
	&& + \frac{n(\vect{r}) \, \gradn \cdot \nabla \left( \frac{\gradn \cdot \gradd}{\left| \gradn \right|} \right)}{8 \left(3 \pi^2\right) n(\vect{r})^4} \\[2ex]
	&& - \frac{3 \, \gradn \cdot \left[ \nabla \left| \gradn \right| \right] \, \delta(\vect{r}-\vect{r'})}{8 \left(3 \pi^2\right) n(\vect{r})^4} \nonumber
\label{eq:Definition_du_dn}
\\[3ex]
\funcder{t(\vect{r})}{n(\vect{r'})} = && \,\, \frac{n(\vect{r}) \, \nabla^2 \delta(\vect{r}-\vect{r'}) - \left[ \nabla^2 n(\vect{r}) \right] \cdot \frac{5}{3} \, \delta(\vect{r}-\vect{r'})}{4 \left(3 \pi^2\right)^{\frac{2}{3}} n(\vect{r})^{\frac{8}{3}}}
\label{eq:Definition_dt_dn}
\end{eqnarray}
Altogether these equations yield the functional derivative of $B(\vect{r})$ in terms of the functional derivatives of $\funcder{s(\vect{r})}{n(\vect{r'})}$, $\funcder{u(\vect{r})}{n(\vect{r'})}$ and $\funcder{t(\vect{r})}{n(\vect{r'})}$:
\begin{eqnarray}
\funcder{B(\vect{r})}{n(\vect{r'})} = - F'(s) \biggl[ && \frac{3}{4} \; \frac{1}{s} \; \funcder{t(\vect{r})}{n(\vect{r'})} - \frac{3}{4} \; \frac{t}{s^2} \; \funcder{s(\vect{r})}{n(\vect{r'})} \nonumber\\
	&& - \frac{3}{4} \; \frac{1}{s^2} \; \funcder{u(\vect{r})}{n(\vect{r'})} + 2 \cdot \frac{3}{4} \; \frac{u}{s^3} \; \funcder{s(\vect{r})}{n(\vect{r'})} \biggr] \nonumber\\[1ex]
	- F''(s) \biggl[ && \frac{3}{4} \; \frac{t}{s} \; \funcder{s(\vect{r})}{n(\vect{r'})} - 2 \cdot \frac{3}{4} \; \frac{u}{s^2} \; \funcder{s(\vect{r})}{n(\vect{r'})} \nonumber\\
	&& + \frac{3}{4} \; \frac{1}{s} \; \funcder{u(\vect{r})}{n(\vect{r'})} - s \; \funcder{s(\vect{r})}{n(\vect{r'})} \biggr] \nonumber\\[1ex]
	- F'''(s) \biggl[ && \frac{3}{4} \; \frac{u}{s} - s^2 \biggr] \funcder{s(\vect{r})}{n(\vect{r'})} 
\label{eq:dB_dn_final_form}
\end{eqnarray}
In none of the Sternheimer eqs.~(\ref{eq:SternPhi+}),~(\ref{eq:SternPhi-}),~(\ref{eq:SternVH+}),~(\ref{eq:SternVxc+}) and~(\ref{eq:Sternn+}) the kernel is needed explicitly standalone. The only point where it enters the formalism is by setting up the exchange-correlation potential response via eq.~(\ref{eq:SternVxc+}). As the GGA form~(\ref{equ:GGA_form_energy}) is an approximation for the exchange energy and is used in the adiabatic approximation
\begin{equation*}
\f{x}^{\text{\tiny SL}}(\vect{r},\vect{r'},\bar{\omega}) = \f{x}^{\text{\tiny SL}}(\vect{r},\vect{r'})
\end{equation*}
throughout this manuscript,
eq.~(\ref{eq:SternVxc+}) together with eq.~(\ref{eq:Kernel_SL_GGA}) becomes
\begin{eqnarray}
V_{\mathrm{x},\text{\tiny SL}}^{+}(\vect{r}) = && \int\mathrm{d}^3 r' \; n^{+}(\vect{r'}) \, \f{x}^{\text{\tiny SL}}(\vect{r},\vect{r'}) = \nonumber\\[2ex]
= && \frac{4}{9} \, A_{\text{x}} \, n(\vect{r})^{-\frac{2}{3}} \, B(\vect{r}) \, n^{+}(\vect{r}) \label{eq:Potential_response_SL}\\
&& + \frac{4}{3} \, A_{\text{x}} \, n(\vect{r})^{\frac{1}{3}} \, \int\mathrm{d}^3 r' \; n^{+}(\vect{r'}) \, \funcder{B(\vect{r})}{n(\vect{r'})}. \nonumber
\end{eqnarray}
In the occuring integral, the integration variable is $\vect{r'}$ and the only dependences on $\vect{r'}$ in $\funcder{B(\vect{r})}{n(\vect{r'})} $ are buried in the $\delta$-functions of $\funcder{s(\vect{r})}{n(\vect{r'})}$, $\funcder{u(\vect{r})}{n(\vect{r'})}$ and $\funcder{t(\vect{r})}{n(\vect{r'})}$. Thus, the $\vect{r'}$ integration in eq.~(\ref{eq:Potential_response_SL}) comes down to integrals of the form
\begin{equation}
\zeta^{+}(\vect{r}) := \int\mathrm{d}^3 r' \; n^{+}(\vect{r'}) \, \funcder{\zeta(\vect{r})}{n(\vect{r'})} ,
\label{eq:Remaining_integrals_zeta}
\end{equation}
where $\zeta$ is $s$, $u$ or $t$, respectively. These integrals are given by:
\begin{eqnarray}
s^{+}(\vect{r}) = && \frac{n(\vect{r}) \, \gradn \cdot \nabla n^{+}(\vect{r}) - \frac{4}{3} \, \left| \gradn \right|^2 \, n^{+}(\vect{r})}{2 \left(3 \pi^2 \right)^{\frac{1}{3}} n(\vect{r})^{\frac{7}{3}} \, \left| \gradn \right|} \nonumber\\\label{eq:s+}\\[3ex]
u^{+}(\vect{r}) = && \,\, \frac{n(\vect{r}) \, \nabla |\gradn| \cdot \nabla n^{+}(\vect{r})}{8 \left(3 \pi^2\right) n(\vect{r})^4} \nonumber\\[2ex]
	&& + \frac{n(\vect{r}) \, \gradn \cdot \nabla \left( \frac{\gradn \cdot \nabla n^{+}(\vect{r})}{\left| \gradn \right|} \right)}{8 \left(3 \pi^2\right) n(\vect{r})^4} \label{eq:u+}\\[2ex]
	&& - \frac{3 \, \gradn \cdot \left[ \nabla \left| \gradn \right| \right] \, n^{+}(\vect{r})}{8 \left(3 \pi^2\right) n(\vect{r})^4} \nonumber\\[3ex]
t^{+}(\vect{r}) = && \frac{n(\vect{r}) \, \nabla^2 n^{+}(\vect{r}) - \left[ \nabla^2 n(\vect{r}) \right] \cdot \frac{5}{3} \, n^{+}(\vect{r})}{4 \left(3 \pi^2\right)^{\frac{2}{3}} n(\vect{r})^{\frac{8}{3}}}
\label{eq:t+}
\end{eqnarray}
At this point the exchange potential response of eq.~(\ref{eq:Potential_response_SL}) is fully determined and can be expressed in terms of $s^{+}(\vect{r})$, $u^{+}(\vect{r})$ and $t^{+}(\vect{r})$, which yields:
\begin{eqnarray}
V_{\mathrm{x},\text{\tiny SL}}^{+}(\vect{r}) = \frac{4}{9} A_{\text{x}} n(\vect{r})^{-\frac{2}{3}} && B(\vect{r}) n^{+}(\vect{r}) \nonumber\\[2ex]
+ \frac{4}{3} A_{\text{x}} n(\vect{r})^{\frac{1}{3}} && \int\mathrm{d}^3 r' n^{+}(\vect{r'}) \funcder{B(\vect{r})}{n(\vect{r'})} = \label{eq:Potential_response_SL_final_form}\\[2ex]
= \frac{4}{9} A_{\text{x}} n(\vect{r})^{-\frac{2}{3}} B(\vect{r}) && n^{+}(\vect{r}) + \frac{4}{3} A_{\text{x}} n(\vect{r})^{\frac{1}{3}} 
I \nonumber \hspace*{3mm} \mathrm{and} \\[3ex]
I= - F'(s) \biggl[ && \frac{3}{4} \; \frac{1}{s} \; t^{+}(\vect{r}) - \frac{3}{4} \; \frac{t}{s^2} \; s^{+}(\vect{r}) \nonumber\\
	&& - \frac{3}{4} \; \frac{1}{s^2} \; u^{+}(\vect{r}) + 2 \cdot \frac{3}{4} \; \frac{u}{s^3} \; s^{+}(\vect{r}) \biggr] \nonumber\\[1ex]
- F''(s) \biggl[ && \frac{3}{4} \; \frac{t}{s} \; s^{+}(\vect{r}) - 2 \cdot \frac{3}{4} \; \frac{u}{s^2} \; s^{+}(\vect{r}) \nonumber\\
	&& + \frac{3}{4} \; \frac{1}{s} \; u^{+}(\vect{r}) - s \; s^{+}(\vect{r}) \biggr] \label{eq:I-II-III_final_form}\\[1ex]
- F'''(s) \biggl[ && \frac{3}{4} \; \frac{u}{s} \; s^{+}(\vect{r}) - s^2 \; s^{+}(\vect{r}) \biggr] \nonumber
\end{eqnarray}
This is the potential response for a spin-independent calculation. 
The spin-scaling relation 
\begin{equation}
\f{x}^{\text{\tiny SL},\sigma , \tau}[n_{\uparrow},n_{\downarrow}](\vect{r},\vect{r'}) = 2 \; \f{x}^{\text{\tiny SL}}[2 n_{\sigma}](\vect{r},\vect{r'}) \; \delta_{\sigma \tau}
\label{eq:Spin_scaling_SL_kernel}
\end{equation}
for the x kernel then leads to
the spin-dependent potential response
\begin{eqnarray}
V_{\mathrm{x},\text{\tiny SL}}^{+,\sigma}(\vect{r}) = \; && \;\; \sum_{\tau = \uparrow,\downarrow} \int\mathrm{d}^3 r' \; n_{\tau}^{+}(\vect{r'}) \, \f{x}^{\text{\tiny SL},\sigma , \tau}[n_{\uparrow},n_{\downarrow}](\vect{r},\vect{r'}) = \nonumber\\[2ex]
= && \;\;\; 2 \int\mathrm{d}^3 r' \; n_{\sigma}^{+}(\vect{r'}) \; \f{x}^{\text{\tiny SL}}[2 n_{\sigma}](\vect{r},\vect{r'}) 
\label{eq:SL_potential_response_spin_dependent}
\end{eqnarray}
in the adiabatic approximation. For implementing a given GGA, it only remains to compute the first, second and third derivatives of the exchange enhancement factor, i.e., $ F'(s), F''(s), F'''(s)$.

\subsection{Deriving the hydrogen atom density response from the Sternheimer equations}
\label{app:1s2pstern}

We start from the Sternheimer eqs.~(\ref{eq:SternPhi+}) and~(\ref{eq:SternPhi-}), which for an exact calculation of the hydrogen atom read
\begin{align}
\begin{split}
\lbrack h - \epsilon_{1 \mathrm{s}} - \hbar\bar{\omega} \rbrack \vert \phi^{+} \rangle = & - \hat{Q} \, V_{\mathrm{ext}}^{+} \, \vert \varphi_{1 \mathrm{s}} \rangle = - V_{\mathrm{ext}}^{+} \, \vert \varphi_{1 \mathrm{s}} \rangle \\
\lbrack h - \epsilon_{1 \mathrm{s}} + \hbar\bar{\omega} \rbrack \vert \phi^{-} \rangle = & - \hat{Q} \, V_{\mathrm{ext}}^{+} \, \vert \varphi_{1 \mathrm{s}} \rangle = - V_{\mathrm{ext}}^{+} \, \vert \varphi_{1 \mathrm{s}} \rangle 
\end{split}
\label{equ:One_electron_Sternheimer_equations_Phi_+_Phi_-}
\end{align}
where $\vert \phi^{+} \rangle$ and $\vert \phi^{-} \rangle$ are the orbital responses of an orbital starting its propagation in the hydrogen atom $1 \mathrm{s}$ ground-state orbital $\vert \varphi_{1 \mathrm{s}} \rangle$, $\epsilon_{1 \mathrm{s}}$ is the eigenenergy of the hydrogen atom $1 \mathrm{s}$ ground-state orbital and $h$ is the ground-state Hamiltonian of the hydrogen atom.\\
The projector $\hat{Q} = 1 - \vert \varphi_{1 \mathrm{s}} \rangle \langle \varphi_{1 \mathrm{s}} \vert$ is of no effect in this case, as $\langle \varphi_{1 \mathrm{s}} \vert V_{\mathrm{ext}}^{+} \vert \varphi_{1 \mathrm{s}} \rangle = 0$. This is because $\vert \varphi_{1 \mathrm{s}} \rangle$ is an even function with respect to its spatial coordinates, but $V_{\mathrm{ext}}^{+}$ is a linear function with regard to its spatial coordinates according to eq.~(\ref{eq:Def_Vext+}) and thus is spatially odd.

The calculation for the hydrogen atom is of course a spin-dependent one. However, as only one spin channel is occupied, and as there is no preference for either of the two possibilities, the spin index is omitted in this section. \\
$\vert \phi^{+} \rangle$ and $\vert \phi^{-} \rangle$ are orthogonal to $\vert \varphi_{1 \mathrm{s}} \rangle$~\cite{fabian} which can easily be verified by projecting $\langle \varphi_{1 \mathrm{s}} \vert$ onto eqs.~(\ref{eq:SternPhi+}) and~(\ref{eq:SternPhi-}). 
Hence, $\vert \phi^{+} \rangle$ and $\vert \phi^{-} \rangle$ can be expanded in terms of the unoccupied ground-state orbitals $\vert \varphi_j \rangle$ with $j > 1$ as $\vert \varphi_1 \rangle := \vert \varphi_{1 \mathrm{s}} \rangle$:
\begin{align}
\begin{split}
\vert \phi^{+} \rangle = \sum\limits_{j = 2}^{\infty} c_{j}^{+} \vert \varphi_{j} \rangle \\
\vert \phi^{-} \rangle = \sum\limits_{j = 2}^{\infty} c_{j}^{-} \vert \varphi_{j} \rangle
\end{split}
\label{equ:Expansion_Phi_+_Phi_-_one_electron}
\end{align}
Inserting this into the Sternheimer eqs.~(\ref{equ:One_electron_Sternheimer_equations_Phi_+_Phi_-}) and projecting $\langle \varphi_{i} \vert$ onto them yields
\begin{align}
\langle \varphi_{i} \vert \lbrack h - \epsilon_{1 \mathrm{s}} - \hbar\bar{\omega} \rbrack \vert \phi^{+} \rangle = & \sum\limits_{j = 2}^{\infty} \left( \epsilon_{j} - \epsilon_{1 \mathrm{s}} - \hbar\bar{\omega} \right) c_{j}^{+} \, \langle \varphi_{i} \vert \varphi_{j} \rangle \notag\\
= & - \langle \varphi_{i} \vert V_{\mathrm{ext}}^{+} \,\vert \varphi_{1 \mathrm{s}} \rangle \\[2ex]
\langle \varphi_{i} \vert \lbrack h - \epsilon_{1 \mathrm{s}} + \hbar\bar{\omega} \rbrack \vert \phi^{-} \rangle = & \sum\limits_{j = 2}^{\infty} \left( \epsilon_{j} - \epsilon_{1 \mathrm{s}} + \hbar\bar{\omega} \right) c_{j}^{-} \, \langle \varphi_{i} \vert \varphi_{j} \rangle \notag\\
= & - \langle \varphi_{i} \vert V_{\mathrm{ext}}^{+} \,\vert \varphi_{1 \mathrm{s}} \rangle \, ,
\end{align}
where $\epsilon_{j}$ is the corresponding hydrogen ground-state energy eigenvalue of $\vert \varphi_{j} \rangle$. From this the coefficients $c_{i}^{\pm}$ can be determined as
\begin{align}
c_{i}^{\pm} = \frac{- \langle \varphi_{i} \vert V_{\mathrm{ext}}^{+} \,\vert \varphi_{1 \mathrm{s}} \rangle}{\epsilon_{i} - \epsilon_{1 \mathrm{s}} \mp \hbar\bar{\omega}} \, ,
\end{align}
which by inserting these coefficients into eq.~(\ref{equ:Expansion_Phi_+_Phi_-_one_electron}) results in
\begin{align}
\vert \phi^{+} \rangle = - \sum\limits_{i = 2}^{\infty} \frac{\langle \varphi_{i} \vert V_{\mathrm{ext}}^{+} \,\vert \varphi_{1 \mathrm{s}} \rangle}{\epsilon_{i} - \epsilon_{1 \mathrm{s}} - \hbar\bar{\omega}} \, \vert \varphi_{i} \rangle 
\label{equ:One_electron_Phi_+_expansion}\\[1ex]
\vert \phi^{-} \rangle = - \sum\limits_{i = 2}^{\infty} \frac{\langle \varphi_{i} \vert V_{\mathrm{ext}}^{+} \,\vert \varphi_{1 \mathrm{s}} \rangle}{\epsilon_{i} - \epsilon_{1 \mathrm{s}} + \hbar\bar{\omega}} \, \vert \varphi_{i} \rangle
\label{equ:One_electron_Phi_-_expansion}
\end{align}
For the further derivation the matrix element 
\begin{equation*}
\langle \varphi_{i} \vert V_{\mathrm{ext}}^{+} \vert \varphi_{1 \mathrm{s}} \rangle \stackrel{(\ref{eq:Def_Vext+})}{=} e E \, \langle \varphi_{i} \vert x \vert \varphi_{1 \mathrm{s}} \rangle
\end{equation*}
has to be calculated, where the external electric field points in the \mbox{$x$-direction} ($\vect{E} = E \, \vect{e}_x$). In order to evaluate this, each of the three parts $\varphi_{i}$, $x$ and $\varphi_{1 \mathrm{s}}$, respectively, can be expressed by spherical harmonics. Ref.~\cite{Cohen_Tannoudji} calculates such integrals of three spherical harmonics, from which the dipole selection rules can be derived. One finds:
\begin{equation}
\langle \varphi_{i} \vert x \vert \varphi_{1 \mathrm{s}} \rangle = \langle \varphi_{n {\mathrm{p}_x}} \vert x \vert \varphi_{1 \mathrm{s}} \rangle \cdot \delta_{i,n {\mathrm{p}_x}}
\end{equation}
With this and eqs.~(\ref{equ:One_electron_Phi_+_expansion}) and~(\ref{equ:One_electron_Phi_-_expansion}) the density response can be expanded in terms of the unoccupied ground-state orbitals according to eq.~(\ref{eq:Sternn+}). With real-valued ground-state orbitals and the definition \mbox{$\hbar\omega_{1i} := \epsilon_{i} - \epsilon_{1 \mathrm{s}}$} one arrives at:
\begin{align}
\begin{split}
n^{+}(\vect{r},\bar{\omega}) & = \varphi_{1 \mathrm{s}}(\vect{r}) \lbrack \phi^{+}(\vect{r},\bar{\omega}) + \phi^{-}(\vect{r},\bar{\omega}) \rbrack = \\[2ex]
& = - \sum\limits_{n=2}^{\infty} \, \frac{eE}{\hbar} \, \langle \varphi_{n {\mathrm{p}_x}} \vert x \vert \varphi_{1 \mathrm{s}} \rangle \, \varphi_{1 \mathrm{s}}(\vect{r}) \, \varphi_{n {\mathrm{p}_x}}(\vect{r}) \\
& \qquad\qquad \times \left[ \underbrace{\frac{1}{\omega_{1n} - \bar{\omega}} + \frac{1}{\omega_{1n} + \bar{\omega}}}_{=: \, C} \right]
\end{split}
\label{equ:n_+_one_electron}
\end{align}
with
\begin{align}
\begin{split}
\Re(C) =& \frac{2 \, \omega_{1n} \, [(\omega_{1n}^2 - \omega^2) + \eta^2]}{[(\omega_{1n} - \omega)^2 + \eta^2][(\omega_{1n} + \omega)^2 + \eta^2]} \\[2ex]
\Im(C) =& \frac{4 \, \omega_{1n} \, \omega \, \eta}{[(\omega_{1n} - \omega)^2 + \eta^2][(\omega_{1n} + \omega)^2 + \eta^2]} ]
\end{split}
\end{align}\\
As already mentioned in section~\ref{sec:Linear_response_TDDFT_calculations}, the parameter $\eta$ is introduced to model the switch-on process of the external perturbation.
In the adiabatic limit of $\eta \rightarrow 0$, the density response for $\omega \neq \omega_{1n}$ becomes 
\begin{align}
n^{+}(\vect{r},\omega) = - \sum\limits_{n=2}^{\infty} \, \frac{e E}{ \hbar} \, & \langle \varphi_{n {\mathrm{p}_x}} \vert x \vert \varphi_{1 \mathrm{s}} \rangle \, \varphi_{1 \mathrm{s}}(\vect{r}) \, \varphi_{n {\mathrm{p}_x}}(\vect{r}) \notag\\
& \times \frac{2 \, \omega_{1n}}{(\omega_{1n} - \omega) \, (\omega_{1n} + \omega)} .
\label{equ:One_electron_n_+_eta_0}
\end{align}
In line with section's~\ref{sec:Asymptotics_of_the_potential_response_of_semi_local_exchange_approximations} objective to evaluate the potential response for the $1 \mathrm{s} \rightarrow 2 \mathrm{p}_x$ excitation, eq.~(\ref{equ:One_electron_n_+_eta_0}) has to be considered in the limit $\omega \rightarrow \omega_{12}$. In this case the term for $n = 2$ dominates all other contributions.
Thus, this yields
\begin{equation}
n^{+}(\vect{r},\omega \rightarrow \omega_{12}) \propto \varphi_{1 \mathrm{s}}(\vect{r}) \, \varphi_{2 {\mathrm{p}_x}}(\vect{r}) 
\label{equ:One_electron_n_+_prop_to_phi_1s_phi_2p}
\end{equation}
for the density response of the $1 \mathrm{s} \rightarrow 2 \mathrm{p}_x$ excitation of the hydrogen atom.
As the interest in section~\ref{sec:Asymptotics_of_the_potential_response_of_semi_local_exchange_approximations} lies only in the spatial dependence of the investigated quantities, we use
\begin{equation}
n^{+}(\vect{r}) = \varphi_{1 \mathrm{s}}(\vect{r}) \, \varphi_{2 {\mathrm{p}_x}}(\vect{r}),
\end{equation}
i.e., drop the proportionality factors.
Inserting the explicit, analytic forms of $\varphi_{1 \mathrm{s}}(\vect{r})$ and $\varphi_{2 {\mathrm{p}_x}}(\vect{r})$~\cite{Cohen_Tannoudji}, eqs.~(\ref{equ:One_electron_density_explicit_form}) and (\ref{equ:One_electron_density_response_explicit_form}) follow.

\clearpage

\bibliographystyle{epj}
\bibliography{ak13responserefs}

\end{document}